\documentstyle[12pt,aaspp4]{article}

\def\<{{\langle}}
\def\>{{\rangle}}
\def\cm{{\rm cm}}
\def\G{{\rm G}}
\def\K{{\rm K}}

\def\Myr{{\rm Myr}}
\def\msun{{\rm\,M_\odot}}

\def\kms{{\rm km\, s^{-1}}}
\def\pc{{\rm pc}}
\def\kpc{{\rm kpc}}
\def\va{v_{\rm A}}

\def\cs{c_{\rm s}}

\def\tg{t_{\rm g}}

\def\nj{n_{\rm J}}

\def\alf{Alfv\'en }

\def\nh2{n_{H_2}}

\def\rb{{\bar{\rho}}}

\def\ekinit{E_{K,\ init}}
\def\sigv{\sigma_v}

\def\simgt{\lower.5ex\hbox{$\; \buildrel > \over \sim \;$}}
\def\simlt{\lower.5ex\hbox{$\; \buildrel < \over \sim \;$}}

\lefthead{Ostriker et al}
\righthead{  }

\begin{document}

\title{Kinetic and Structural Evolution of Self-gravitating, Magnetized Clouds:
2.5-Dimensional Simulations of Decaying Turbulence}

\author{Eve C. Ostriker$^1$, Charles F. Gammie$^{2,3}$, and James M. Stone$^1$}
\affil{$^1$Department of Astronomy, University of Maryland \\
College Park, MD 20742-2421}
\affil{$^2$Isaac Newton Institute for Mathematical Sciences\\
20 Clarkson Rd., Cambridge, CB3 0EH, UK}
\affil{$^3$Harvard-Smithsonian Center for Astrophysics, MS-51 \\
60 Garden St., Cambridge, MA 02138}

\begin{abstract} 

The molecular component of the Galaxy is comprised of turbulent,
magnetized clouds, many of which are self-gravitating and form stars.
To develop an understanding of how these clouds' kinetic and
structural evolution may depend on their level of turbulence, mean
magnetization, and degree of self-gravity, we perform a survey of
direct numerical MHD simulations in which three parameters are
independently varied.  Our simulations consist of solutions to the
time-dependent MHD equations on a two-dimensional grid with periodic
boundary conditions; an additional ``half'' dimension is also
incorporated as dependent variables in the third Cartesian direction.
Two of our survey parameters -- the mean magnetization parameter
$\beta\equiv c_{sound}^2/v_{Alfven}^2$ and the Jeans number $n_J\equiv
L_{cloud}/L_{Jeans}$ -- allow us to model clouds which either meet or
fail conditions for magneto-Jeans stability and magnetic criticality.
Our third survey parameter -- the sonic Mach number ${\cal M}\equiv
\sigma_{velocity}/c_{sound}$ -- allows us to initiate turbulence of
either sub- or super- Alfv\'enic amplitude; we employ an isothermal
equation of state throughout.  We evaluate the times for each cloud
model to become gravitationally bound, and measure each model's
kinetic energy loss over the fluid flow crossing time.  We compare the
evolution of density and magnetic field structural morphology, and
quantify the differences in the density contrast generated by internal
stresses, for models of differing mean magnetization.  We find that
the values of $\beta$ and $n_J$, but not the initial Mach number $\cal
M$, determine the time for cloud gravitational binding and collapse:
for mean cloud density $n_{H_2}=100 \cm^{-3}$, unmagnetized models
collapse after $\sim 5\Myr$, magnetically supercritical models
generally collapse after $5-10 \Myr$ (although the smallest
magneto-Jeans stable clouds survive gravitational collapse until
$t\sim 15\Myr$), while magnetically subcritical clouds remain
uncollapsed over the entire simulations; these cloud collapse times
scale with the mean density as $t_g\propto n_{H_2}^{-1/2}$.  We find,
contrary to some previous expectations, less than a factor of two
difference between turbulent decay times for models with varying
magnetic field strength; the maximum decay time, for $B\sim 14\mu\G$
and $n_{H_2}=100 \cm^{-3}$, is 1.4 flow crossing times
$t_{cross}=L/\sigma_{velocity}$ (or $8\Myr$ for typical GMC
parameters).  In all models, we find turbulent amplification in the
magnetic field strength up to at least the level $\beta_{pert}\equiv
c_{sound}^2/\delta v_{Alfven}^2=0.1$, with the turbulent magnetic
energy between 25-60\% of the turbulent kinetic energy after one flow
crossing time.  We find that for non-self-gravitating stages of
evolution, when clouds have ${\cal M}=5-10$, the mass-averaged density
contrast magnitudes $\langle\log(\rho/\bar\rho)\rangle$ are in the
range $0.2-0.5$, with the contrast increasing both toward low- and
high- $\beta$.  Although our conclusions about density statistics may
be affected by our isothermal assumption, we note that only the more
strongly-magnetized models appear consistent with estimates of
clump/interclump density contrasts inferred in Galactic GMCs.

\end{abstract}


\section{Introduction}

All present-day star formation in our Galaxy is observed to take place
in cold molecular clouds.  These clouds are highly turbulent (with
sonic Mach numbers 10 or more for the most massive clouds, e.g. 
\cite{bli93}), and appear
to be quite inhomogeneous based on both molecular line observations in
a range of tracers (e.g. \cite{fal92}, \cite{fal98}), 
and stellar extinction maps (\cite{lad94}, \cite{lad98}).  
Because magnetic field measurements are difficult to obtain, the distributions
of magnetic field strengths and directions within clouds are less well
characterized than other cloud properties; the mean field strengths
are probably no less than a few $\mu\G$, and no more than a few tens
of $\mu\G$ (e.g. \cite{hei93}, \cite{cru93}, \cite{tro96}, \cite{cru98}), 
and the spatial power spectrum of the magnetic field 
is probably dominated by components with close to 
the largest scales possible, within structures of a given density regime 
(e.g. \cite{goo94}, \cite{goo95}, \cite{sch98})
Given the extremely low thermal pressure in molecular clouds, 
magnetic stresses are expected to be at least as
important as gas pressure in governing cloud evolution, and may be
much more so.  In particular, since sufficiently strong mean magnetic
fields would render the internal motions sub-Alfv\'enic and the cloud
self-gravity subcritical, it has long been argued that
magnetization may significantly inhibit turbulent dissipation and
prevent gravitational collapse (\cite{aro75}, \cite{mou76}, 
\cite{shu87}, \cite{mck93}).

The nonlinear amplitudes and structural irregularity of the turbulence
seen in molecular clouds require time-dependent numerical simulations
in order to model their evolution theoretically.  This is the second
in a series of papers reporting on the results of numerical
experiments designed to characterize how magnetic fields at a range of
strengths affect the structure and evolution of cold, turbulent,
self-gravitating clouds.  In Paper I (\cite{gam96}), we outlined (see
also \cite{ost97}) the observational situation, the theoretical
background on the nature of ideal turbulent flows and applications to
cold Galactic clouds, and the issues that can be addressed
by numerical simulations; we then presented results of the first
extensive survey (in ``1 2/3D'' slab geometry, i.e. with one
independent spatial variable, but all three components of dependent
spatial variables) of model cloud evolution in the ideal MHD limit.
The survey considered clouds both with and without gravity, and either
with an input spectrum of turbulence that decayed over time, or with
an ongoing excitation of turbulence through velocity perturbations
introduced at spatial scales smaller than the simulation box.  Among
the conclusions of Paper I was the finding that, when sufficient
initial energy in the form of transverse disturbances is input to a
cloud model and then evolved, gravitational collapse along the mean
field can be suppressed for times $>\tg$ (see eq. \ref{tgdef} below).
Paper I also reported on the
large density contrasts, and correspondingly small filling factors,
which arise in freely-decaying turbulent MHD flows (e.g. a typical
model having 50\% of the mass at density larger than three times the
mean, and 10\% of the mass at nine times greater than the mean).
Because these conclusions could potentially be affected by the
restricted geometry of the slab-symmetric models -- where no spatial variations
perpendicular to the mean field are possible -- we 
re-examine these issues in the present paper with a less restricted 
geometry. 

Recent numerical work by other groups has addressed a variety of related  
questions.  \cite{elm97} has found that hierarchical density structure 
arises in 1D MHD cloud models which are subject to mechanical forcing at their
boundaries.  \cite{pas95} and \cite{vaz96}  have modeled cloud formation and
evolution in the ISM on $\sim\kpc$ scales, incorporating a number
of physical effects (Galactic rotation and shear, heating from 
localized star formation, and cooling) in addition to magnetic fields.  
\cite{pad97a} and \cite{mac98} have simulated the decay of Mach-5 
turbulence in 3D cloud models with weak ($c_s^2/v_A^2=2,1$, respectively) and 
moderate ($c_s^2/v_A^2=0.02,0.04$, respectively) magnetic fields, and 
find less than a factor 2 difference in the kinetic energy decay for the
two cases.

Because the matter distributions in clouds should eventually influence
star formation and the IMF, numerical modelers have taken a particular
interest in analyzing the density structure of their simulated clouds.
In particular, there has been considerable recent attention devoted to
the shape of the density distribution function obtained in simulations
of compressible turbulence. Based on a variety of non-self-gravitating
simulations at varied Mach numbers, with/without turbulent forcing,
and using a variety of polytropic indices $\gamma$, two different groups
(\cite{pad97b}, \cite{nor98}; \cite{vaz94}, \cite{sca98}, and
\cite{pas98}) have concurred in finding that log-normal density
distributions are expected when $\gamma=1$, and the latter group finds
skewed log-normal/power law distributions are expected when $\gamma\ne
1$.  These groups differ considerably, however, in their predictions of the 
scaling of the mean density contrast magnitude with Mach number when $\gamma=1$
(see \S 3.3).

In this paper, we extend our previous slab-geometry survey (Paper I)
by performing a survey in 2 1/2 dimensions\footnote{ i.e. two spatial
independent variable, but three spatial components to all dependent
variables; this allows spatial variations both parallel and
perpendicular to the mean magnetic field.} of the evolution of
turbulence in model clouds with a range of sizes, Mach numbers, and
magnetizations.  In particular, since incompressible motions (i.e.
nonlinear Alfv\'en waves) have often been considered as a
weakly-dissipative effective ``turbulent'' pressure that can prevent
cloud collapse over several free-fall times (see Paper I), we focus on
the evolution of self-gravitating model clouds in which an initial,
incompressible turbulent spectrum is initiated, but no subsequent
small-scale energy is injected.  Thus, although our models are highly
idealized and should not be thought of as representing real clouds, our 
results are a useful guide for understanding how cloud dynamics would proceed
in the absence of potential turbulent inputs, many of which have been 
attributed to the onset of 
internal star formation (e.g. \cite{nor80}, \cite{mie94}, \cite{pas95}).  
The plan of the paper is as follows:
\S 2 describes our numerical experiments, 
including the computational code we use (\S2.1), the
definitions of relevant dimensionless parameters (\S2.2), and our choices of
parameters for the survey (\S2.3). In \S3, we present our results on
the evolution of cloud energies (\S3.1), structural appearance (\S3.2), 
and density distributions (\S3.3) in our models.  We summarize our results and
discuss the principal implications of our work for the evolution and 
structure of observed clouds in \S4.

\section{Simulations}

\subsection{Numerical Method}

To create the model cloud evolutions presented in this paper, we
integrate the equations of ideal, compressible, self-gravitating MHD
using a variant of the ZEUS algorithm developed by
\cite{sto92a,sto92b}.  ZEUS is an operator-split, finite difference
method on a staggered mesh.  The magnetic field is evolved using
``constrained transport'' (\cite{eva88}), which guarantees that
$\nabla \cdot {\bf B} = 0$ to machine precision.  The transverse
components of the magnetic field are evolved using the ``method of
characteristics'' which guarantees accurate propagation of Alfv\'enic
disturbances.  The ZEUS algorithm has been extensively tested and used
in a wide variety of astrophysical applications.  In the present
models, dissipation occurs on small scales in part due to purely
numerical effects associated with discretizing the MHD equations (i.e.
from zone-to-zone averaging of velocity, magnetic field, or density
variations), and in part due to an artificial viscosity introduced to
capture shocks.  We do not include ambipolar diffusion (which is
important on small scales in GMCs), or other explicit resistivity or
viscosity.

Our implementation of ZEUS is ``2 1/2 dimensional'' (2.5 D).  This
peculiar nomenclature means that there are two independent spatial
variables ($x,y$), but that all (three) components of the velocity and 
magnetic field vectors are still evolved as dependent variables.  In this
sense, there are two fully dynamical dimensions and a third dimension in which
the flow is symmetric. The boundary conditions in all models are periodic.

The gravitational potential is determined from $\nabla^2 \phi = 4\pi
G(\rho - \bar{\rho})$; the mean density appears because of the periodic
boundary conditions.  We solve for the gravitational potential using the
Fourier method.  The gravitational kernel is
\begin{equation}\label{poisseq}
\phi_{\bf k} = 2\pi G\rho_{\bf k} \left(
        (1 - \cos(k_x \Delta x))/\Delta x^2
        + (1 - \cos(k_y \Delta y))/\Delta y^2 \right)^{-1}.
\end{equation}
This ensures that the discrete representation of Poisson's equation
is satisfied.  

We adopt, for all models, an isothermal equation of state, meaning
that $p = c_s^2 \rho$, where $c_s^2$ is a constant in both space and
time.  Thus we do not integrate an internal energy equation.  A
non-adiabatic assumption is a physically reasonable approximation for
the higher-density ($n_{H_2}\sim 10^2- 10^4\cm^{-3}$) portions within
molecular clouds because the cooling time is very short, leading to a
range of observed temperatures $10-20\K$ over molecular regions not
directly associated with high-luminosity star formation (e.g.
\cite{gol87} and references therein).  If radiative processes are
sufficiently rapid compared to dynamical times, then an effective
polytropic index $\gamma$ may be found by balancing heating and
cooling rates locally (\cite{sca98}).  The value of the effective
$\gamma$ increases from $\gamma<1$ to $\gamma \simgt 1$ near
$n_{H_2}\sim 10^3 \cm^{-3}$, with the increase due to the inefficiency
of cooling in optically thick regions (radiative trapping).  The
details of the transition in this effective $\gamma$ depend on the
relation of the density at a given velocity to the column density at
that velocity (cf.  \cite{gll78}, \cite{deJ80}, \cite{neu95}); this
relation however remains statistically uncertain (as well as spatially
multivalued) in a turbulent medium.  Since no single choice of an
effective $\gamma$ applies to all density regimes of interest in the
interior of a GMC, it would be highly desirable to perform numerical
calculations in which both thermodynamics and MHD equations are
evolved in detail.  This would be a formidable calculation to perform
realistically, because the molecular cooling lines are optically
thick; to date, no group has done hydrodynamic calculations with even
model heating and cooling included, in the regime appropriate for
GMCs.  At this stage, when the effects of magnetic fields on turbulent
dynamics are not understood, we view the choice $\gamma=1$ as an
acceptable compromise because most of the gas evolves to
higher-than-average densities for which the temperature likely varies
by less than a factor $\sim 2$, even if a strictly isothermal
effective $\gamma$ applies only to $n_{H_2}\simgt 10^3 \cm^{-3}$.  The
isothermal assumption is less accurate for the low-density
``interclump'' regions that form as a result of turbulent dynamics.
We comment further on the implications of our choice of $\gamma$ in \S
3.3.

For all the simulations, we choose the initial velocity field to be locally
divergence-free (i.e. the individual Fourier components are constrained 
to satisfy ${\bf v_k} \cdot {\bf k} = 0$), and isotropic in direction.  
The velocity amplitude 
is drawn from a Gaussian distribution with power spectrum 
$\langle |{\bf v_k}|^2 \rangle\propto k^{-3}$, 
for $k=|k_x\hat x + k_y \hat y|$ 
and $2\pi/L\le k_i \le 128 (2\pi/L)$.  The initial kinetic energy therefore 
obeys $E=\Sigma_{k_x,k_y} E_{\bf k}$ where 
$E_{\bf k}=(1/2) v_{\bf k}^2 \propto k^{-3}$, 
corresponding to a relationship 
$\sigma_v(R)\propto R^{1/2}$ between 
size scale $R$ and the velocity dispersion $\sigma_v$ averaged over that 
scale;  we choose this initial spectrum to match 
the linewidth-size relation for 
observed self-gravitating, molecular clouds on a variety of scales.
For comparison, we note that a Kolmogorov spectrum in 2D has 
$\langle |{\bf v_k}|^2 \rangle\propto k^{-8/3}$, so that  
$\sigma_v(R)\propto R^{1/3}$.

Initially, the density for all models is taken to be uniform; all
inhomogeneities are generated self-consistently from divergences in
the velocity field created by internal pressure forces, gravity, and
magnetic stresses (since ${\nabla\cdot \bf v}=0$ initially).  The
initial magnetic field is also taken to be uniform, pointing in the
$\hat x$ direction (right-left in the model snapshots shown).  Notice
that the uniformity of $\bf B$ and $\rho$ in the initial conditions
implies that the mass-to-flux ratio along each field line is constant.
In snapshots shown, the symmetric direction $\hat z$ points out of
page.

We have tested our implementation of the ZEUS algorithm using a variety
of test problems: linear waves, standard advection tests, shock tubes
and the parametric instability of circular polarized Alfv\'en waves.
Of particular interest for what follows is whether the simulations are
converged at our standard resolution, $256^2$ zones.  We have tested
this by evolving an initial Gaussian random
velocity field, as in all our decay experiments, while holding the amplitude
and phase of the field's Fourier components fixed as the resolution
is increased from $32^2$ to $512^2$ zones. 
The convergence test models use $n_J=2.5$, $E_{K, init}=50$, 
and $\beta=0.02$ (see \S 2.2 for definitions).  We find a secular 
decrease in the ratios $E^{2n}/E^n$ of turbulent 
energies ($\equiv$ perturbed magnetic and kinetic energy) 
at successive resolutions; at time $0.1t_s$ these ratios decrease
from $E^{64}/E^{32}=1.19$ to $E^{128}/E^{64}=1.11$ to $E^{256}/E^{128}=1.08$
to $E^{512}/E^{256}=1.06$.  We also find only a 5\%
difference in the gravitational binding time (see \S3.1) between the 
$256^2$ and $512^2$ models.  Thus, at a resolution of $256^2$ zones, 
the evolution is satisfactorily converged.

\subsection{Scalings and definitions}

The fundamental dimensional units for our simulations are the length $L$ of
the box edge, the mean density $\rb=M/L^3$ of matter in the box,
and the (isothermal) sound speed $c_s=\sqrt{kT/\mu}$.  Because
self-gravity is crucial in the problems under study, it is useful to
describe the size of the simulated region relative to the minimum
Jeans-unstable wavelength $L_J\equiv \cs (\pi/\G \rb)^{1/2}$ at the
cloud's mean density and temperature.  For each simulation we
therefore define the size via the Jeans number $n_J\equiv L/L_J$;
using $c_s=0.19\, \kms (T/10\K)^{1/2}$ for $\mu=2.4m_p$ the corresponding
physical size is
\begin{equation}\label{LJdef}
L= n_J L_J=n_J\times 1.9\,\pc
        \left({T\over{10\K}} \right)^{1/2}
        \left({n_{H_2}\over{10^2\cm^{-3}}} \right)^{-1/2}.
\end{equation}
With the parameter values $n_J=2,3,4$ used in the present simulations, 
the box edge would respectively correspond to
approximately $4, 6, {\rm or}\ 8\pc$ at the mean conditions in a
molecular cloud ($n_{H_2}\sim 100$), a factor 3 smaller for conditions
within a cloud clump ($n_{H_2}\sim 10^3$), or a factor 10 smaller for
conditions in a dense core ($n_{H_2}\sim 10^4$).

Time in our numerical experiments is measured either in 
units of the sound-crossing time $t_s\equiv L/c_s$ for the simulated region, or in
terms of a characteristic gravitational contraction time 
$t_g\equiv L_J/c_s=(\pi/G\rb)^{1/2}$.\footnote{For reference, the free-fall
collapse times 
for a cold sheet, cylinder, or sphere are, respectively $0.25t_g$, $0.28t_g$, 
and $0.31t_g$.}
The corresponding physical time depends on the mean density according to    
\begin{equation}\label{tgdef}
t_g= t_s/n_J= 9.9\, \Myr \left({n_{H_2}\over{10^2\cm^{-3}}} \right)^{-1/2}.
\end{equation}
Thus for $n_J=2,3,$ or 4, the interval $0.1 t_s$ would respectively
correspond to approximately $2, 3, {\rm or}\ 4\ \Myr$ at the typical
mean density in a molecular cloud, and a factor 3 or 10 shorter at
clump or core densities; the interval $0.1t_g\sim 1 \Myr$ at typical
mean GMC conditions.

The total mass within a cube with faces the same size as 
the simulated region corresponds to $M=n_J^3 M_J$, where
the Jeans mass at the mean conditions is
\begin{equation}
M_J\equiv \rb L_J^3 =48\,\msun\left({n_{H_2}\over{10^2\cm^{-3}}} \right)^{-1/2}
        \left({T\over{10\K}} \right)^{3/2}.
\end{equation}

At the beginning of all the simulations, the magnetic field is uniform with 
strength $B_0$ and points in the $\hat x$ direction. The level of the mean
magnetic field in each simulation is characterized by the ratio 
\begin{equation}\label{betadef}
\beta\equiv {\cs^2\over \va^2} = {\cs^2\over B_0^2/(4\pi\rb)}=
0.021\left({T\over{10\K}} \right)
\left({n_{H_2}\over{10^2\cm^{-3}}} \right)
\left({B_0 \over 10 \mu\G}\right)^{-2}
\end{equation}
corresponding to $B_0=1.4\mu\G \beta^{-1/2} \left({T\over{10\K}} \right)^{1/2}
\left({n_{H_2}\over{10^2\cm^{-3}}} \right)^{1/2}$;
decreasing $\beta$ corresponds to increasing the importance of magnetic fields
to the dynamics.   For the values $\beta=0.01,\ 0.1, 1.0$ used in the present
set of simulations, the mean magnetic field is $B_0=14,\ 4.4,$ and $1.4\mu\G$ 
respectively, for fiducial values $T=10\K$ and $n_{H_2}=100\cm^{-3}$.

Uniform clouds are unstable to compressions transverse to the mean magnetic 
field (``magneto-Jeans unstable'') when the magnetosonic wave
crossing time exceeds $t_g$ (\cite{cha53}); for $\va>>\cs$ this occurs
when $n_J>\beta^{-1/2}$, corresponding to having the cross-field
column density $N_{H_2}> 4.1\times 10^{21} \cm^{-2} (B_0/10\mu\G)$ .
A closely-related criterion is used to evaluate a more stringent stability
condition, as follows:
If a uniform cubic cloud with size $L=n_J L_J$ condenses along the mean 
field to make a cold ``pancake'' (with central density 
$\rho_0=\Sigma^2\pi\G/(2\cs^2)=(\pi\nj)^2\bar\rho/2$),
the pancake will have ratio of
surface density to magnetic field $\Sigma/B_0=(n_J \beta^{1/2})/(2
G^{1/2})$.  This magnetized pancake is unstable to fragmentation when
$\Sigma/B_0>1/(2\pi G^{1/2})$
\footnote{The same instability criterion
applies when considering either the ratio $\Sigma/B_0$ for a uniform
magnetized sheet or the central mass-to-magnetic-flux ratio of a
magnetized cloud (\cite{tom88}).} (corresponding to 
$N_{H_2}> 1.3 \times 10^{21} \cm^{-2} (B_0/10\mu\G)$),  
so that the corresponding
instability criterion in terms of the original cloud's parameters  
is $n_J>\beta^{-1/2}\pi^{-1}$. 
Clouds with $n_J>\beta^{-1/2}\pi^{-1}$  are termed ``supercritical;'' clouds
with the inequality reversed are termed ``subcritical.''  We note that the 
gravitational plus magnetic energy of the ``pancake'' is still negative
provided $6/(\pi\nj)^4<\beta$.

In terms of the fundamental units, all energies in the problem are given in
units of $\rb\cs^2 L^2$; because the $\hat z$ direction is symmetric, this 
is the energy per unit length $dz$.  For a cubic cloud, the corresponding 
total energy unit would be $\rb\cs^2 L^3$.  Where the kinetic energy is 
given, it corresponds to 
\begin{equation}
E_K={1\over 2} \int_{-L/2}^{L/2}\int_{-L/2}^{L/2} dx dy 
(v_x^2 + v_y^2 + v_z^2)\rho ;
\end{equation}
the square of the Mach number ${\cal M} \equiv \sigma_v/c_s$ (where $\sigma_v$ 
represents the total velocity dispersion) is therefore
${\cal M}^2 =(\sigma_v/c_s)^2= 2 [E_K/(\bar\rho L^2 c_s^2)]$, i.e. 
twice the normalized kinetic energy.
The total magnetic energy corresponds to 
\begin{equation}\label{ebdef}
E_B={1\over 8\pi} \int_{-L/2}^{L/2}\int_{-L/2}^{L/2} dx dy 
(B_x^2+B_y^2+B_z^2);
\end{equation}
for the energy in the perturbed magnetic field, $\delta E_B$, we subtract out 
the mean-field energy $1/(2\beta)$ from equation (\ref{ebdef}).  
Because we assume periodic boundary conditions, the mean magnetic field 
(corresponding to the $k=0$ Fourier component of $\bf B$), and hence the
mean-field energy, remains unchanged over the course of any simulation.  
The gravitational energy is
\begin{equation}
E_G={1\over 2} \int_{-L/2}^{L/2}\int_{-L/2}^{L/2} dx dy (\phi \rho)
\end{equation}
where the gravitational potential is computed from the 
modified Poisson equation in  periodic boundary conditions (\S1.1). 
We note that because equation ({\ref{poisseq}) implies 
$\phi/\cs^2\propto L^2 G \rb/\cs^2$, the scaled gravitational energy will
obey $E_G/(\rb L^2 \cs^2)\propto n_J^2$. 

\subsection{Survey Model Parameters}

As described in \S1, observed clouds present a range of properties, with 
some more tightly constrained than others.  From a theoretical point of view,
the observed 
bulk cloud properties of size, mean magnetization, and velocity dispersion
can be described in terms of the dimensionless parameters $n_J$, 
$\beta$, and ${\cal M}^2=2 [E_K/(\bar\rho L^2 c_s^2)]$ 
introduced in \S2.2.  The well-known 
correlations between cloud properties (\cite{lar81}) 
can be converted into 
relationships between these parameters, too.  For example, linewidth/size and
density/size scalings of the form $\sigma_v\propto L^{1/2}$ and 
$\bar\rho \propto L^{-1}$ reported either for whole GMCs (\cite{sol87}) or 
for clumps within GMC's (\cite{bal87}) 
can be combined to show that cloud kinetic energy ($E_K\propto {\cal M}^2$) 
and gravitational energy ($E_G\propto n_J^2$) scale together; in terms of
the Mach number $\cal M$ and Jeans number $n_J$, 
the observed relationship can be expressed quantitatively as 
${\cal M}=\sigma_v/c_s=1.7 n_J$.  
With this observed scaling for sonic Mach number, 
the corresponding \alf\ Mach number would satisfy  
${\cal M}_A \equiv \sigma_v/ v_A = 1.7 n_J \beta^{1/2}$.  As a consequence,
these observed clouds would be magneto-Jeans unstable 
($n_J \beta^{1/2} >1$) when ${\cal M}_A >1.7$, and magnetically supercritical 
($n_J \beta^{1/2} >1/\pi$) when ${\cal M}_A >0.54$.  
We can also define a fluid 
flow crossing time $t_{cross}\equiv L/\sigv$;  
using the observed scalings, this implies $\tg=1.7 t_{cross}$. 

In modeling the evolution of decaying turbulence in idealized magnetized, 
self-gravitating clouds, our survey of numerical experiments brackets the
range of observed properties wherever possible.  
The survey covers a variety of different initial conditions by varying three 
parameters independently:  we set the size of the box to $n_J=2, 3,$ or 4 
Jeans lengths (see eq.\ref{LJdef}), we set the ratio 
$c_s^2/v_A^2 \equiv\beta=0.01, 0.1, 1,$ or $10^6$ (see eq. \ref{betadef}), 
and we set the initial kinetic energy to 
$E_{K, init}/(\rb\cs^2 L^2)= 25, 50,$ or $100$, corresponding to  
initial sonic Mach numbers of ${\cal M} =7.1, 10,$ or $14$.
To allow for the decay of turbulent energy, we have 
set the range of initial kinetic energies slightly higher than estimates of
observed cloud kinetic energies in our range of $n_J$.  
The initial \alf\ Mach number 
${\cal M}_A\equiv 
\langle v^2/\va^2\rangle^{1/2}=\left(2\beta E_K\right)^{1/2}
={\cal M}\beta^{1/2}$ 
is 7.1, 10, or 14 for the $\beta=1$ models; 2.2, 3.2, or 4.5 for the 
$\beta=0.1$ models; and 0.71, 1, or 1.4 for the $\beta=0.01$ models.
Thus, all of the models with $\beta=10^6, 1, 0.1$ have initially 
significantly super-Alfv\'enic flow, while the models 
with $\beta=0.01$ have initially sub-Alfv\'enic or slightly 
super-Alfv\'enic flow.  
For $\beta=1$, all of the model clouds would be unstable by the magneto-Jeans
criterion (if initially unperturbed), and also supercritical.   
For $\beta=0.1$, the $\nj=4$ models would be magneto-Jeans 
unstable and supercritical,  while the $\nj=3, 2$ models would be 
magneto-Jeans stable but supercritical.  For $\beta=0.01$, 
the $\nj=4$ models would be magneto-Jeans stable but supercritical,  while
the $\nj=3,2$ models would be magneto-Jeans stable and subcritical.
The set of parameters used
for the simulation suite is given in the left three columns of Table 
\ref{tbl-1}.

\section{Results}

\subsection{Overall Cloud Energetics}

Figure 1 shows the evolution of the total ``perturbed'' energy 
$E_{tot}\equiv E_K+ \delta E_B + E_G$ for all of the cloud models over the
course of the simulations. 
From Figure 1, it is clear that low-magnetization
cloud models become gravitationally-bound -- defined as reaching
$E_{tot}<0$ -- at significantly earlier times $t_{bind}$ than 
high-magnetization models, and that the models with the same values of 
$\beta$ and $\nj$ (i.e. same mean magnetic field and size) which become
bound all do so at close to the same time, 
regardless of the initial level of turbulent 
kinetic energy.  In Table 1, we list the time $t_{bind}$ 
(or lower limit) at which the model clouds become bound. The 
nearly-unmagnetized ($\beta=10^6$)
cloud models all become bound at approximately the same time, $0.5 t_g$, 
regardless of the initial level of kinetic energy or cloud size (i.e. $\nj$).
This is only slightly larger than the time ($\approx0.3 t_g$) that a cold, 
initially-quiescent cloud would take to collapse; thus, turbulent energy 
does not in itself do much toward preventing cloud collapse.
The $\beta=1$ models become bound at times 
between $0.5-1.1 t_g$, with the larger ($\nj=4$) clouds collapsing before the
smaller ($\nj=2$) ones.  Thus, weakly-magnetized 
model clouds which have \alf Mach numbers larger than one, and are 
magneto-Jeans unstable, collapse within a factor of two of the time for 
collapse of completely unmagnetized clouds.  The $\beta=0.1$ model clouds
all have initial \alf Mach numbers $>1$, as well, and are all supercritical.
Of these, 
the $\nj=4$ models, which are magneto-Jeans unstable, collapse at approximately
the same time ($\sim 0.6 t_g$) as the nearly-unmagnetized ($\beta=10^6$) 
and weakly-magnetized $\beta=1$ models;
the $\nj=3$ models (which are just barely magneto-Jeans stable) 
remain unbound  
until $0.7-1.0 t_g$, while the magneto-Jeans stable (but supercritical)
$\nj=2$ models evolve to $1.5 t_g$ before undergoing rapid gravitational 
collapse.  For $\beta=0.01$, all cloud models with $\nj=2,3$ (which are 
subcritical) remain unbound
up to times at least $t_g$, with the models with higher initial $E_K$ lasting
until the simulations were halted at $1.5t_g$; the supercritical 
$\nj=4$ clouds lose energy more quickly and collapse at $\sim 0.8\tg$.  

Consistent with expectations, 
the gravitational energy of the subcritical cloud models ($\beta=0.01$ and
$\nj=2,3$) never exceed the energy of a self-gravitating thin sheet
$E_{G, sheet}\approx -[(\nj^2\pi^2/6)-1]$, over
the course of the simulations ($t>t_g$; see Table 1).  Since the time for an 
initially-quiescent, cold cloud to collapse to a sheet is just $0.25 t_g$, 
it is clear that the time-dependent magnetic field perturbations have delayed
the collapse along the mean field, in subcritical clouds.  
For the other models that do become 
bound, the gravitational energy exceeds the energy of a self-gravitating sheet
at approximately the same times ($t_{condense}$) 
as the times when they become bound ($t_{bind}$) (see Table 1).  

For all models, Figure 1 shows that there is an initial phase of rapid
kinetic energy loss, with the dissipation rate higher in the models
with the larger initial kinetic energy and weaker mean magnetic
fields.  The time dependence of the kinetic energy dissipation is
isolated in Figure 2, which shows the evolution of $E_K$ for all of
the cloud models on a log-log scale.  Over the period before the
matter becomes strongly self-gravitating, after a transient phase
(during which much of the initial energy is lost) the kinetic energy
declines approximately proportional to $t^{-0.3}$ in the $\beta=0.01,
\ 0.1$ models, and proportional to $t^{-0.6}$ in the $\beta=1$ models.
The kinetic energy evolution in the $\beta=10^6$ models does not have
a significant period of power-law decay.\footnote{In recent work by
ourselves (see \cite{sto98}) and another group (\cite{mac98}),
numerical simulations of decaying magnetized and unmagnetized
turbulence in 3D have shown that the kinetic energy decline during
the power-law phase is a steeper function of $t$.  We believe that
the difference with the present 2.5D simulations arises because
motions in the symmetry direction ($\hat z$) are dissipated less
easily.  Because of the difference between 2.5D and 3D, we regard
the present dissipation rates as lower limits; however, we note that
the quantitative differences in the total magnitude of dissipation between
2.5D and 3D simulations performed to date are not large (see \S4.1)}

Since the simulations pesented in this paper all describe models of decaying,
rather than quasi-steady-state, turbulence, there is no single number which
can be used to characterize the turbulent decay rate for each model.  However,
a useful way to quantitatively 
characterize the differences in the evolution of the kinetic
energy for model clouds of different mean magnetization and size is to compare
the total kinetic energy after one crossing time with the initial kinetic 
energy.  We define a characteristic crossing time for the flow as 
$t_{cross}=L/(2 E_{K,\ init})^{1/2}$; for $E_{K,\ init}=25,\ 50,\ 100$, 
$t_{cross}/t_s=0.1414,\ 0.1,\ 0.0707$, respectively.  In Table 1, we list 
the values of the
ratio $E_K(t_{cross})/E_{K,\ init}$ for each model cloud.  For the $\beta=0.01$
models, this ratio is in the range $0.42-0.48$, with the value depending
more on the initial kinetic energy than on $\nj$ because self-gravity 
does not become important for any of these model clouds within $t_{cross}$.  
For the $\beta=0.1$ models, all of the models which do not become strongly
self-gravitating within $t_{cross}$ have $E_K(t_{cross})/E_{K,\ init}$ 
smaller than the corresponding values for the $\beta=0.01$ models;  the range
is $E_K(t_{cross})/E_{K,\ init}=0.32-0.46$.  All of the $\beta=10^6$ models 
have smaller values of $E_K(t_{cross})/E_{K,\ init}$ (in the range 
$0.26-0.32$ for uncollapsed models) than the corresponding
models with $\beta=1$, and most of the $\beta=1$ models have smaller values for
this ratio ($0.33-0.39$) than the $\beta=0.1$ models.  Although the trends in 
the relative levels of dissipation are as expected, with the lower-$\beta$ 
cloud models experiencing less dissipation than the higher-$\beta$
models, there is less than a factor two enhancement of kinetic energy 
at times $t_{cross}$ in the
former compared to the latter.  Thus, magnetic fields do reduce dissipation,
but when the \alf Mach number is of order 0.5 or larger, as for the present
set of models (and as expected from estimates of field 
strengths in self-gravitating molecular clouds), the reduction in dissipation 
is not large.

In Figure 3, we show the evolution of the perturbed magnetic energy 
$\delta E_B$ for all models.  Comparison with Figure 2 shows that 
the perturbed magnetic energy in the 
$\beta=0.01,0.1,\ {\rm and}\ 1$ models typically 
increases until its value is $1/3-2/3$ that of the kinetic energy,
and then both decline together.  The time
to reach the peak magnetic energy increases with increasing $\beta$ and 
decreasing $\ekinit$, and lies in the range $0.1-0.25 t_A$ where the \alf
crossing time $t_A\equiv L/\va=t_s \beta^{1/2}$.  Overall, the peak values of
$\delta E_B$ decrease with increasing $\beta$, and depend only weakly on 
$\nj$ for any value of $\ekinit$.  A fit to the peak value of $\delta E_B$ 
good to $\pm 30\%$ is 
\begin{equation}\label{ebmaxfit}
\delta E_{B,\ max} =0.3 \ekinit^{0.8} \beta^{-0.18};
\end{equation} 
this fit includes only those models with $\beta=0.01, 0.1, 1$ for 
which $\delta E_B$ reaches a maximum and then declines.
In Table 2, we list the
ratio $\delta E_B/E_K$ after one crossing time.  For $\beta=0.01, 0.1, 
{\rm and}\ 1$, this ratio varies between 0.25-0.62; for $\beta=10^{-6}$
this ratio is just $4-10 \times 10^{-6}$.
Because the peak value
of $\delta E_B$, the time at which the peak is attained, and the rate of
kinetic energy dissipation all depend on $\beta$, the ratio 
$\delta E_B/E_K(t_{cross})$ varies with $\beta$ for fixed $\ekinit$;  the
variation with $\nj$ at fixed $\beta$ and $\ekinit$ is more moderate.  
Defining $\beta_{pert}\equiv \cs^2/\delta \va^2= 1/(2 \delta E_B)$, 
we note that for all the models with $\beta=0.01$, 0.1, and 1, the minimum 
value of $\beta_{pert}$ is 0.1 or smaller;  
thus, there is dynamical growth of the 
perturbed magnetic field due to turbulence 
even when the initial mean field is relatively weak.  The amount of growth of 
the perturbed magnetic field likely depends on the degree of helicity in the
initial velocity field (cf. \cite{men81}, \cite{haw96}), which is close to 
maximal for our choice of a divergence-free initial velocity field.

\subsection{Evolution of Cloud Structure}

To illustrate the growth of density structure in model clouds, Figure 4 
shows density snapshots for the model with $\beta=0.01$, $\nj=3$, and 
$\ekinit=50$ at time intervals $0.1t_g$ ($\sim 1\Myr$).  
This model is magneto-Jeans stable, and marginally subcritical.
The early structure, consisting of many small clumps and interconnected
filaments, is characteristic of the early appearance of all the models.  Over
time, the small structures agglomerate into larger structures as kinetic energy
is lost to dissipation 
and gravity does its work.  This general process occurs in the other
models, as well.  However, the late-time structural evolution of the models
differs qualitatively depending on the values of $\beta$ and $\nj$.  
In subcritical models, 
over time gravity causes matter to slide along the magnetic fields
and collect into larger filaments (actually the projections of sheets in this 
2D geometry) which lie preferentially perpendicular to the mean field.  
Figure 4 shows how these sheets may continue to oscillate in time.  In 
supercritical
models, on the other hand, the self-gravity of the condensations that form
is sufficient to overwhelm magnetic stresses, leading to the growth of 
rounder clumps and eventual gravitational collapse.

Figure 5 compares the development of structure in three cloud models
which differ only in the magnitude of their mean magnetic field --
having $\beta=0.01, 0.1,$ and 1.  The $\beta=0.01$ model is
magneto-Jeans stable and (marginally) subcritical; the $\beta=0.1$
model is magneto-Jeans stable but supercritical; the $\beta=1$ model
is magneto-Jeans unstable and supercritical.  The model clouds all
have the same size $\nj=3$, and all have the same value of
$\ekinit=50$ (and, in fact, identical initial spatial 
velocity perturbations).  In the $\beta=0.01$ model, the magnetic field lines
remain close to their original, parallel configuration; early-time
large oscillations of the matter clumps and filaments bend and
compress the field locally.  At the final time shown ($t=0.8 t_g$),
the matter density remains less than $\rho/\rb=100$ everywhere.  In
the $\beta=0.1$ model, the magnetic field is less rigid and hence
there are much larger initial departures from parallel lines.  Local
regions with density $\log(\rho/\rb)>2$ appear in the $t/t_g=0.5$
snapshot (these contain 3.6\% of the total mass), and grow in volume
and mass so that at $t/t_g=0.8$, 18\% of the mass is at
$\log(\rho/\rb)>2$.  The magnetic field becomes more uniform as
kinetic energy is lost, but there remain large departures from
straight lines in the high-density regions.  In the $\beta=1$
simulation, the magnetic fields immediately become strongly bent and
stretched, and by $t/t_g=0.8$ has even undergone noticable reconnection 
in the high-density region (due to numerical effects).  The matter becomes
strongly clumped, with 8 \% of the mass at $\log(\rho/\rb)>2$ at
$t/t_g=0.5$, and 62 \%(55\%) of the mass at $\log(\rho/\rb)>2
(3)$ at $t/t_g=0.8$.  In this last frame, a single zone contains
47\% of the total mass; because the density in this zone is high
enough to make the Jeans length smaller than the grid scale, the
simulation is not well-resolved at this point.  The $\beta=10^6$
simulations (not shown) are also initially filamentary, and gradually
coagulate due to gravity.  The
late-time structures consist of rotationally-supported condensations
(with trailing spiral arms); this sort of structure does not develop in
the lower-$\beta$ models because magnetic stresses redistribute
angular momentum away from gravitationally-collapsing regions.

\subsection{Statistics of Matter Distributions}


As described in \S 3.2, the overall structure in our model clouds
evolves in time due to the combined action of gravity, Reynolds
stresses, and magnetic stresses.  At the same time as the overall
structure evolves, the detailed distributions of matter in the models
evolves as well.  These distributions can be characterised in a
variety of ways.  We find it informative to consider two related
distributions: the fraction of volume $dV/V_{tot}$, and the fraction
of mass $dM/M_{tot}=\rho dV/(\bar\rho V_{tot})$, as a function of the 
logarithmic density enhancement/decrement $\log(\rho/\rb)$.  In Figure 6,
we show how these volume and mass distributions appear for the cloud
model snapshots portrayed in Figure 5.  The values of the volume-averaged and 
mass-averaged density logarithms 
\begin{equation}\label{volavg}
\langle \log(\rho/\rb)\rangle_V \equiv \sum 
{\log(\rho/\rb) dV(\rho)\over V_{tot}}
\end{equation}
and 
\begin{equation}\label{massavg}
\langle \log(\rho/\rb)\rangle_M \equiv \sum 
{\log(\rho/\rb) \rho dV(\rho)\over \bar\rho V_{tot}}
\end{equation}
are also indicated, for each distribution.  

A number of common
features are apparent for all the models shown in Figure 6;
these trends apply to our other models, as well: (a) The mass distributions
are offset toward larger density than the volume distributions; this
simply means that matter is clumped, with the mass-averaged density
logarithm larger than the volume-averaged density logarithm.  Because
most of the volume is at lower-than-average density, while most of the
mass is at higher-than-average density, $\langle
\log(\rho/\rb)\rangle_V$ is negative and $\langle
\log(\rho/\rb)\rangle_M$ is positive.  (b) At earlier times, the
distributions are very roughly log-normal in shape, while large tails
evolve in time.  (c) The values of $\langle
\log(\rho/\rb)\rangle_V$ and $\langle
\log(\rho/\rb)\rangle_M$ are nearly equal in magnitude, at early
times.  It is easy to show that for a lognormal distribution, this
will be true by definition.  (d) Over time, the mass and volume
distributions become more separated, for any given model.  This is due
to the action of gravity increasing the clumping of matter.  
(e) At the earliest
time, before self-gravity becomes important, the $\beta=0.01$  model has
greater density contrast than the $\beta=0.1, 1$ models.

Given the strong temporal evolution in any of our models, no single
distribution characterises its density structure.  Because
the density distribution in a real cloud is a fundamental property, however,
it is important to investigate in our model clouds how the density
distribution depends on the input parameters.  If a clear correlation
exists between the density distribution and input parameters in our
model clouds, then the density contrast in real clouds (estimated from 
molecular mapping), and the more detailed distributions of column 
density (determined from extinction mapping), can potentially be used to
constrain other cloud properties such as the mean magnetic field strength.
While the limitations of our isothermal equation of state may affect our
conclusions in detail, we still consider it important to investigate
how turbulence and magnetic fields alone impact the density structure
that develops. 

To isolate the effect of the magnetic field in influencing the density 
structure
in our model clouds, we have chosen to analyze the models early in their 
evolution, before the effects of gravity become significant.  Since the
velocity dispersion -- corresponding to the kinetic energy in our models --
is the primary direct observable in real clouds, we select instants in each
model cloud's evolution when the kinetic energy has a fixed value; because 
$E_K$ decays at different rates for different $\beta$, these analysis 
times differ for the different cloud models.  To make quantitative comparisons
among the models, we  
evaluate the volume- and mass- averaged density logarithms
at times in each model when the kinetic energy has decayed to half of its
initial value.  Thus, for models with initial $E_{K, init}=100, 50, 25$, the
respective values of the Mach number at the time of density structure 
measurement are ${\cal M}=10, 7.1, 5$. 
The computed values for 
$\langle \log(\rho/\rb)\rangle_V$ and 
$\langle \log(\rho/\rb)\rangle_M$
are listed in the last two columns of Table 1.  In a schematic 
picture of clouds as consisting of denser clumps embedded 
within a more tenuous interclump medium, we may
think of $\langle \log(\rho/\rb)\rangle_M$ as describing the 
magnitude of the positive
density contrast in the ``clump'' component, and 
$\langle \log(\rho/\rb)\rangle_V$ as describing the magnitude of the
negative density contrast in the ``interclump medium''.  

Based on the computed values of $\langle \log(\rho/\rb)\rangle_V$ and 
$\langle \log(\rho/\rb)\rangle_M$ tabulated, we can make a number of 
observations.  These remarks apply to the set of 27 models 
with $\beta=0.01,0.1,1$;  since 
the unmagnetized models have unrealistically low field strength, and 
generally become self-gravitating earlier, we do not consider them here.
At fixed $E_K$ and $\beta$, we find the variation of the
the density contrast with $\nj$ is (except for the 
$\beta=0.01$, $\ekinit=0.25$ models) less than $0.15$;  thus, these variations
are not principally due to gravitational effects (as desired).  We also 
find that for fixed $\beta$, the density contrasts 
$\langle \log(\rho/\rb)\rangle_{V, M}$  generally 
increase logarithmically with increasing $E_K$.  Finally, we find
that at fixed $E_K$, all except for two models with $\beta=0.01$ have larger
density contrasts than the corresponding $\beta=0.1,1$ models at all $\nj$.
In general, we find that the variation of density contrast with Mach number 
at fixed $\beta$ is smaller than the variation of density contrast with
$\beta$ at fixed Mach number, for the parameter range studied.

In Figure 7, we plot the values of the clump and interclump density contrasts
{\it vs} the total magnetic energy density, so that the models with different
mean $\beta$ fall in three separate groups.  Within each group, increasing
(perturbed) magnetic energy correlates with increasing kinetic energy, 
accounting for the increase shown in the density 
contrast toward larger total $E_B$.  The 
$\beta=0.1$ points lie quite close to the lines 
$|\langle \log(\rho/\rb)\rangle_{V,M}|=C \log(B^2/(8\pi\bar\rho c_s^2))$ 
with $C=0.2-0.25$, and
the $\beta=0.01$ points lie near the extension of these lines.  The $\beta=1$
points all lie well above these lines, and typically have density 
contrasts that fall between the $\beta=1$ and $\beta=0.01$ models, at a 
given $E_K$.  

Physically, the decrease in the density contrast in our models in going from
$\beta=1$ to $\beta=0.1$ at fixed $E_K$ can probably be attributed to
a increase in the effective pressure provided by magnetic fields: for
a fixed value of the Mach number $=(2E_K)^{1/2}$, the Alfv\'en Mach
number decreases proportional to $\beta^{1/2}$, and ram-pressure
induced density contrast should decrease as the mean magnetic field strength
increases.  The increase of the density
contrast in going from $\beta=0.1$ to $\beta=0.01$, on the other hand,
is likely the signature of the importance of transverse magnetic
fields in compressing the medium.  At quasiequilibrium 
kinks in a strong magnetic
field, matter is compressed until the central thermal pressure $\rho
c_s^2$ equals the external magnetic pressure $B_\perp^2/8\pi$.  For a nonlinear
transverse wave, $B_\perp^2/4\pi\sim \bar\rho v_\perp^2$, implying 
$\rho/\bar\rho \sim (1/2)(v_\perp/\cs)^2$ at kinks 
(independent of $\beta$).  In 
this situation, the induced density contrast would {\it not} decrease with
increasing mean magnetic field strength. The measured 
tendency for matter to collect more strongly at nulls of the
transverse magnetic field for lower-$\beta$ simulations may in part
owe to the increasing virulence at low $\beta$ of the decay
instability (in which Alfv\'en waves pump the growth of field-aligned
compressive waves and backscattered Alfv\'en waves; e.g.  \cite{sag69},
\cite{gol78}), which may cause the lowest-$\beta$ models to approach the
kinked-field state more rapidly. 

Our results for density contrast can be compared with those reported
by \cite{pad97b}, and those of \cite{pas98} for isothermal models.
The former cites (based on 3D simulations of \cite{nor98}) a result
$-0.5\log(1+{\cal M}^2/4)$ for the value of
$\langle\log(\rho/\bar\rho)\rangle_V$, whereas \cite{pas98} (from 1D
unmagnetized simulations) find the result $-0.22 {\cal M}^2$ for the
same quantity.  The measurements in our Table 1 are made for ${\cal M}=10, 7.1,
5$, so that the \cite{pad97b} formula predicts $\pm 0.7, 0.6, 0.4$, while
the \cite{pas98} formula predicts $\pm 22, 11, 5$ for the values of 
$\langle\log(\rho/\bar\rho)\rangle_{M,V}$.  Because \cite{pad97b} do not
state the value of $\beta$ (or other properties of their simulations), we
cannot make direct comparisons with our results;  nevertheless, the range 
of values $\pm 0.2 -1$ for the density contrast in Table 1 is consistent
with the values they predict.  Our range of values is considerably smaller
than the values predicted by \cite{pas98} for purely 1D turbulence.  We have
also recently found, using forced turbulence simulations in 3D (\cite{sto98}),
comparable values for the density contrast to those obtained in the present
2 1/2 D models.

As a cautionary remark, we note that our conclusions may potentially
be affected by the assumption of isothermality.  \cite{pas98} found,
using 1D simulations, that a power law may develop over a range of
densities larger (smaller) than the mean for polytropic $\gamma<1$
($\gamma>1$); although they do not state how
$\langle\log(\rho/\bar\rho)\rangle_{M,V}$ depend on $\gamma$,
presumably these quantities vary as well.  The power-law regime
increases with increasing $\cal M$ and $|1-\gamma|$.  However, because
the 1D simulations at $\gamma=1$ appear to overestimate the dispersion
and mean density contrast relative to 2.5D and 3D simulations, at any
value of $\cal M$ and $\gamma$ the range of values of
$\log(\rho/\bar\rho)$ over which a power law would occur in
higher-dimension simulations is likely much smaller than those
presented by \cite{pas98}.  In addition, since $\gamma\simgt 1$ ($\gamma<1$)
probably applies for densities larger (smaller) than $\sim$ten
times the mean for molecular cloud conditions (cf. \cite{sca98}), it
is not clear whether a power law would develop within the molecular
component for a substantial portion of a real cloud's internal density
distribution.  \footnote{On the other hand, in the large-scale ISM for which
the effective $\gamma<1$ over a larger range of densities across different
gas phases, a larger power-law regime should develop;  this result has been
found in the analysis by \cite{sca98} of non-polytropic 2D simulations.}
At the time when density statistics are computed in our
models, most of the material lies at $0<\log(\rho/\bar\rho)<1$, so
that for any realistic $\gamma$ the temperature would not have
departed far from the mean value (as is indeed observed in molecular
clouds); in this sense, our isothermal assumption is self-consistent
for the simple mean density contrast statistic which we have measured.
Thus, while calculations incorporating heating and cooling are clearly
warranted, we speculate that at least the values of
$\langle\log(\rho/\bar\rho)\rangle_M$ which would be obtained would
not be far different from those calculated here with an isothermal
assumption.  We discuss potential implications of our results on
density contrast in \S4.3.

\section{Discussion}

\subsection{Summary and comparison with other work}

In this paper, we present the results of a suite of 2.5-dimensional 
simulations following the evolution of idealized 
cloud models with dynamical parameters
comparable to those believed to hold in Galactic molecular clouds.  Our
models, which assume cold, isothermal conditions,
are initiated with random, turbulent velocity fields, and 
include the effects of self-gravity and magnetic forces.  We parameterize the
importance of magnetic fields by $\beta=\cs^2/v_A^2$ 
(see eq. \ref{betadef}), and the importance of self-gravity by the Jeans number
$n_J\equiv L/L_J$ (see eq. \ref{LJdef}).  We follow the 
evolution of cloud models over time as they dissipate kinetic energy in shocks,
and as the clumps, sheets, and filaments formed by the action of magnetic and 
Reynolds stresses coalesce under their own  self-gravity.  

In \S3.1, we compare for different models the times needed for clouds 
to become gravitationally bound, and the times needed to reach a state with 
stronger gravitational binding than a thin sheet.  We find that only the 
models which are magnetically subcritical ($\beta^{1/2} n_J<1/\pi$) survive 
gravitational collapse for times significantly longer than $t=t_g$; in
\S4.2 below we comment on the implications of these results.  We have also 
compared the evolution of kinetic energies, and perturbed magnetic energies, 
among our models.
In examining the amplification of the perturbed magnetic field, we found 
that the peak 
value of $\delta E_B$ attains higher amplitudes, and occurs at earlier times,
for the lower-$\beta$ models (see eq. \ref{ebmaxfit}); at the peak, the value
of $\delta E_B$ is $1/3-2/3$ times the value of $E_K$.  To quantify
the decay of turbulent energy, we compare the kinetic energy 
after a cloud flow crossing time (i.e. an ``eddy turnover time,'' in the
terminology of incompressible turbulence) with the initial
kinetic energy.  We find, for models that have not become strongly 
self-gravitating at this 
time, that the kinetic energy is a fraction 0.26-0.29, 0.33-0.36, 0.32-0.46,
and 0.42-0.48 of the initial value, for $\beta=10^6, 1, 0.1,$ and
$0.01$, respectively, so that 
$\langle E_K\rangle/(\Delta E_K/\Delta t)=0.9-1.4 t_{cross}$.  
Thus, while the models with stronger fields have
lowered dissipation, the effect occurs at a level less than a factor of two,
for realistic field strengths.  
Although the 3D simulations of \cite{mac98} and
\cite{pad97a} have somewhat different parameters, they also find kinetic
energy reduced to 0.3-0.5 of the initial value, after one crossing time, and
differences of no more than a factor two between their low-$\beta$ and 
high-$\beta$ cases.  In recent forced, 3D simulations (\cite{sto98}), we have
found dissipation times $E_W/\dot E$ for the total perturbed energy  
a factor 1-1.75 times smaller than the loss time for half of the  
wave energy for the present 2 1/2 D models; for purely 2D models, the 
dissipation rate is even higher than in 3D.  On the other hand, the 
dissipation in our set of 1D 
models (Paper I) was much less than in the present set, probably because 
the 1D simulations' initial velocity fields were purely transverse, and 
the model symmetry inhibited dissipation.  Since real clouds
are more similar to the 2.5D and 3D models than the 1D models, we
conclude that realistic turbulent dissipation times $E_K/\dot E_K$ 
are not likely to be more than twice the cloud fluid crossing 
times -- i.e., $\simlt 10$ Myr -- for any likely
value of the mean magnetic field. 
Although the hope of substantially lowering turbulent dissipation rates in
molecular clouds by including magnetic fields (\cite{aro75}) may not be 
fully realized, the feedback provided by star formation (e.g. \cite{nor80}) 
on timescales comparable to the turbulent decay (see \S 4.2) may yet 
allow a quasi-steady state to exist (e.g. \cite{mck89}, \cite{pas95}, 
\cite{gam96}).

We present, in \S3.2, snapshots showing density contours, magnetic field 
lines, and the velocity field in cloud models with varying levels of the 
mean magnetic field strength (i.e. $\beta$).  All models show highly
inhomogeneous density structure well before gravity becomes important in 
the dynamics.  However, as gravity becomes important, the matter in 
subcritical clouds collapses into sheets, whereas in supercritical 
clouds it collapses into rounder clumps (and reaches much higher densities).
In \S4.2 below, we summarize our results (based on the analysis in \S3.3)
on the distributions of matter in our models, and comment on implications 
for interpreting the clumpy structure of observed clouds.

\subsection{Implications for cloud support and the initiation of 
star formation}

Observed molecular clouds have column densities in the range
$N_{H_2}=3-10\times 10^{21} \cm^{-2}$ (e.g. \cite{lar81}, \cite{mye88b},
\cite{bli93}).  The corresponding magnetic field strengths required for such 
clouds
to be cross-field (``magneto-Jeans'') stable (see \S2.2) are in the range 
$7-24\mu\G$, within the range of field strengths that are believed to 
prevail on the large scale in molecular clouds (\S1).  However,
typical observed field strengths fall short of the value $23-77\mu\G$
which would be required to render the same whole clouds magnetically 
subcritical; i.e.,
able to resist gravitational fragmentation if all of the cloud's mass
is collected into a sheet, rather than homogeneously filling a volume
with sides comparable to its lateral extent.  We note that projection effects
tend to decrease the observed (line-of-sight) field strength relative to the
3D value, and increase the column density relative to its value along the 
shortest axis, so that observed $B,$ $N_{H_2}$ values might still be 
consistent with marginally critical clouds.

In the simulations
presented in this paper, we show that regardless of the initial
turbulent energy content of our model clouds, only those clouds which
are magnetically subcritical remain uncollapsed throughout the simulations
for times exceeding $1.5t_g$ -- corresponding to $\approx 15$Myr for 
typical conditions.  In fact, the only supercritical clouds that survive
uncollapsed beyond $\tg\sim 10\Myr$  
are relatively small ($\nj=2$, $L\sim 4\pc$) 
and magneto-Jeans stable.
Very low magnetization clouds collapse in $0.5 t_g\sim 5\Myr$; this 
puts a lower limit on the collapse time for supercritical clouds.
If the Milky Way molecular clouds containing most of the mass 
have comparable field strengths to those in observed regions, and column 
densities are at the high end of the range mentioned above, then they
would be supercritical and our simulations would imply that 
they cannot survive for periods longer 
than $5-10\Myr$ without some of part of their interiors becoming strongly
gravitationally bound, and, we expect, beginning to form stars.  Alternatively,
if most of the molecular material is in low column density, subcritical 
clouds, then the dissipation of turbulence incorporated from formation 
would lead to a flattened configuration supported by magnetic fields (if 
there is no turbulent resupply from outside sources), but not dependent on 
turbulent input by stars to avoid further collapse.
This is one of the main conclusions of the present work.

Observationally-based estimates of the time to initiate star formation are
necessarily indirect, and subject to major uncertainty.  Based on 
\cite{wil97}'s analysis of the joint distributions of OB associations and 
GMCs (assuming all molecular material is associated with discrete clouds
with a spectrum of masses $dN/dM\propto M^{-1.6}$), if star formation is 
equally distributed among all clouds, then roughly 80\% of the Galactic 
molecular material would reside in clouds containing one or more OB stars.  If
we take this fraction as representing the portion of a cloud's life during
which it contains an OB association, and use 20 Myr (\cite{bla91}) as the
typical observed lifetime of associations, this implies that associations form
in clouds after 20\% of their 25 Myr lifetimes, 
essentially as soon as our models predict that star formation is
initiated in supercritical clouds (5-10 Myr).  On the other hand, 
even in the largest existing cloud catalog (\cite{sol87}) created for the
inner Galaxy, $\sim 60\%$ of molecular material was not actually assigned to 
discrete clouds, perhaps because it has not been sufficiently warmed internally
by massive stars to stand out above the background (\cite{sol89}).\footnote{
For a fixed number of OB associations, reducing the fraction of clouds 
containing massive stars increases the photoevaporation rate for those
clouds that {\it do} contain associations, making estimates of cloud 
destruction timescales more consistent with observed OB association lifetimes 
(\cite{wil97}).}  If this cold material is in clouds with the same mass
distribution as the warm clouds, we would infer that 60\% of a cloud's lifetime
passes before OB stars form (and it becomes warm).   Under this second 
scenario of unequally-distributed OB stars, typical cloud lifetimes would be 
50 Myr, and a period of $20-25\Myr$ must therefore be
spent in forming exclusively low mass stars, if the clouds are supercritical
and the first stars form after $5-10 \Myr$.  If, on the other hand, much of
the cold molecular material is distributed in small, subcritical clouds that
never make OB associations, then OB star formation in supercritical clouds
(comprising less than half of the total molecular material) 
would again be inferred to commence soon after clouds first begin to have
portions of their interior collapse.   

Based on the above discussion, the range of star formation scenarios
permitted by observations is fairly large at present.  In particular, the
crucial question of the fraction of molecular material residing in sub- and 
super-critical clouds is essentially unknown.  However, the
planned high-resolution Boston University/FCRAO Galactic Ring Survey
in $^{13}$CO and CS (M. Heyer, personal communication) should
considerably constrain the molecular cloud distribution function, and
correspondingly constrain the possible evolutionary scenarios for star
formation and cloud destruction.  This, in turn, should allow us
better to evaluate the conclusions of this paper on the time to
initiate star formation in magnetized GMCs.

\subsection{Implications for clumpy structure of GMCs}
The other finding which we would like to highlight concerns the
distribution of density found in our simulated clouds, and the
potential to use observed density distributions in real clouds as a
diagnostic of the magnetic field strength.  We find, in concurrence
with other recent work on isothermal cloud models, that the shapes of
the density distribution functions before self-gravity becomes
important are roughly log-normal.  Our own analysis of density
distributions presented in \S 3.3, however, primarily focuses not on
the shape of the density distribution function, but instead on the
values of volume-weighted and mass-weighted means of the logarithmic
density, $\langle \log(\rho/\rb)\rangle_V $ and $\langle
\log(\rho/\rb)\rangle_M$.  These represent the magnitudes,
respectively, of the (negative) density contrast in the tenuous
interclump region, and the (positive) density in the cloud clumps,
relative to the mean cloud density.  For a lognormal distribution,
$\langle \log(\rho/\rb)\rangle_V $ and $\langle
\log(\rho/\rb)\rangle_M$ will be equal in magnitude and opposite in
sign; this result holds for most of our models, when analyzed at early
times.  We find that the values of the mean density contrast depend
both on the amplitude of the turbulence (i.e. the value of $E_K$), and
on the strength of the mean magnetic field and its perturbations (i.e.
the value of $E_B$).  Specifically, we find that for a given $\beta$,
the density contrast increases with $E_K$, and that virtually all of
the more strongly-magnetized ($\beta=0.01$) models have larger mean
density contrast than the more weakly-magnetized ($\beta=1,0.1$)
models with the same value of $E_K$.  We also find that the $\beta=1$
models generally have higher density contrast than the $\beta=0.1$
models.  Given the limitations of the current set of simulations, we
did not attempt to formulate a more general theory for the
relationship between the mean density contrast and the values of $E_K$
and $\beta$.  We defer this much-desired goal until the completion of
surveys of 2D and 3D forced, non-self-gravitating MHD turbulence which
are now underway.  With 3D simulations, one can compute column density
distributions, for better comparsion with directly-observed cloud
properties.  We emphasize that our model clouds are highly idealized,
and in particular we employ an isothermal equation of state which may
affect our conclusions (see \S 3.3.).  Still, it is significant
theoretically, and potentially useful astronomically, that the mean
magnetic field strength affects the density distribution.

While follow-up studies are clearly needed, in particular considering 
non-isothermal models, it is 
nevertheless tempting to compare our present results with observations.  
Molecular clouds are often characterized as consisting of clumps embedded in
an interclump medium, with a fraction $\theta_H\equiv M_H/M_{tot}$ 
of the mass, and $f_H=V_H/V_{tot}$ 
of the volume contained in the high density component 
($\rho_H/\bar\rho=\theta_H/f_H$), and the remainder of the mass and volume 
dispersed in a low-density component ($\rho_L/\bar\rho=(1-\theta_H)/(1-f_H)$).
From $^{13}$CO maps of Orion A and the Rosette clouds, 
\cite{bal87} and \cite{wil95} estimate $\theta_H=0.75,0.77$ and 
$f_H=0.1,0.08$, respectively.
With this simple clump/interclump structure, we use
\begin{equation}\label{simpvol}
\langle \log(\rho/\rb)\rangle_V = f_H \log(\rho_H/\bar\rho)
+ (1-f_H)\log(\rho_L/\bar\rho)
\end{equation} 
and
\begin{equation}\label{simpmass}
\langle \log(\rho/\rb)\rangle_M = \theta_H \log(\rho_H/\bar\rho)
+ (1-\theta_H)\log(\rho_L/\bar\rho).
\end{equation} 
to compute
density contrast magnitudes of
$\langle \log(\rho/\rb)\rangle_V=-0.4, -0.5$ and 
$\langle \log(\rho/\rb)\rangle_M=0.5, 0.6$ for the two clouds.
The clumps defined in these observations are not believed to be
self-gravitating; thus, it is fair to compare to the model 
density contrast magnitudes computed in \S3.3.  Of the present models with 
$\beta=1, 0.1, 0.01$ (corresponding to $B_0\sim 1.4, 4.4, 14.4\mu \G$),
Figure 7 shows 
that only the $\beta=0.01$ simulations with Mach number 10
have density contrasts as
large as $\pm 0.5$;  thus, the stronger-field cases seem to be in better 
agreement with observations, when taken at face value.  As argued in \S 3.3,
both $\beta >1$ and $\beta<0.01$ could yield larger density contrast,
although the former (corresponding to sub-$\mu\G$ fields)
is unlikely to exist in the present-day Galaxy.
The implication that $B_0\simgt 15 \mu\G$ 
is far from definitive, but demonstrates the potential to develop
magnetic field diagnostics from other cloud properties such as
the inferred clump/interclump density contrast (which is albeit somewhat 
model-dependent), and the observed distributions of column density 
(obtained from extinction studies, and requiring only an 
$A_V\rightarrow N_{H_2}$ conversion).  We view this as an an important 
direction for future research.

\acknowledgements

We are grateful to B. Elmegreen, J. Ostriker, F. Shu, and the referee
J. Scalo for careful reading and helpful comments on the manuscript of 
this paper, and D. Neufeld for useful discussions on molecular cooling.

This work was supported in part by NASA grant NAG 53840.




\clearpage
\begin{deluxetable}{lrrrrccrr}
\footnotesize
\tablecaption{Evolutionary Characteristics of Model Clouds\label{tbl-1}}
\tablewidth{450pt}
\tablehead{
\colhead{$\beta$} 
& \colhead{$\nj$}
& \colhead{$E_{\rm K, init}$} 
& \colhead{${t_{bind}\over t_g}$\tablenotemark{a}} 
& \colhead{$t_{condense}\over t_g$\tablenotemark{b}} 
& \colhead{$E_K(t_{cr})\over E_{\rm K, init}$\tablenotemark{c}} 
& \colhead{$\delta E_B\over E_K(t_{cr})\tablenotemark{c}$} 
& \colhead{$\langle log(\rho/\rb)\rangle_V\tablenotemark{d}$} 
& \colhead{$\langle log(\rho/\rb)\rangle_M\tablenotemark{e}$} 
}
\startdata
 0.01& 2&  25& $>$1.0&$>$1.0 &0.46 &0.39 & -0.4 & 0.3\nl
 0.01& 2&  50& $>$1.7&$>$1.7 &0.47 &0.31 &-0.5& 0.4\nl
 0.01& 2& 100& $>$1.5&$>$1.5 &0.42 &0.39 &-0.5 &0.5 \nl
 0.01& 3&  25& 0.9&$>$1.0 &0.45 &0.40 &-0.5 &0.5 \nl
 0.01& 3&  50& $>$1.5&$>$1.5 &0.46 &0.31 & -0.4&0.4 \nl
 0.01& 3& 100& $>$1.5&$>$1.5 &0.42 &0.39 & -0.4& 0.4\nl
 0.01& 4&  25& 0.7&0.8 &0.48 &0.33 & -0.7&0.9 \nl
 0.01& 4&  50& 0.9\tablenotemark{\dagger}&$>$0.8 &0.47 &0.32 &-0.5  &0.5 \nl
 0.01& 4& 100& $>$0.8&$>$0.8 &0.42 &0.39 & -0.5& 0.5\nl
  0.1& 2&  25& 1.4&1.2 &0.45 &0.24 & -0.2& 0.2\nl
  0.1& 2&  50& 1.6\tablenotemark{\dagger}&1.5 &0.36 &0.51 & -0.3&0.3 \nl
  0.1& 2& 100& 1.6\tablenotemark{\dagger}&1.4 &0.32 &0.61 & --&-- \nl
  0.1& 3&  25& 0.8\tablenotemark{*}&0.8\tablenotemark{*} &0.46 &0.28 &-0.2&0.3 \nl
  0.1& 3&  50& 1.0&0.9 &0.36 &0.50 &-0.3 &0.3 \nl
  0.1& 3& 100& 0.9\tablenotemark{\dagger}&0.8\tablenotemark{*}  &0.32 &0.62 &-0.3 &0.3 \nl
  0.1& 4&  25& 0.6\tablenotemark{*}&0.6\tablenotemark{*} &0.50 &0.49 & -0.2&0.3 \nl
  0.1& 4&  50& 0.6\tablenotemark{*}&0.6\tablenotemark{*} &0.41 &0.49 &-0.3 &0.3 \nl
  0.1& 4& 100& 0.6\tablenotemark{\dagger}&0.6 &0.32 &0.62 &-0.3 &0.3 \nl
  1& 2&  25& 0.9&0.8 &0.33 &0.54 & -0.2&0.3 \nl
  1& 2&  50& 1.1&0.9 &0.34 &0.38 &-0.3 &0.3 \nl
  1& 2& 100& 1.0\tablenotemark{*}&0.7 &0.36 &0.25 &-- & --\nl
  1& 3&  25& 0.5\tablenotemark{*}&0.5\tablenotemark{*} &0.39 &0.52 &-0.3 & 0.3\nl
  1& 3&  50& 0.7\tablenotemark{*}&0.6\tablenotemark{*} &0.34 &0.39 &-0.3 &0.3 \nl
  1& 3& 100& 0.7\tablenotemark{*}&0.7\tablenotemark{*} &0.35 &0.25 &-0.4 &0.4 \nl
  1& 4&  25& 0.5&0.5 &0.70 &0.52 & -0.4& 0.4\nl
  1& 4&  50& 0.6&0.5 &0.37 &0.40 & -0.4&0.4 \nl
  1& 4& 100& 0.6\tablenotemark{*}&0.6\tablenotemark{*} &0.35 &0.26 &-0.4 &0.4 \nl
  $10^6$& 2&  25& 0.6&0.6 &0.29 &$9\times10^{-6}$ &-0.4 &0.4 \nl
  $10^6$& 2&  50& 0.5&0.5 &0.27 &$5\times10^{-6}$  & -0.4& 0.4\nl
  $10^6$& 3&  25& 0.5&0.6 &0.32 &$9\times10^{-6}$  & -0.5& 0.5\nl
  $10^6$& 3&  50& 0.5&0.6 &0.26 &$6\times10^{-6}$  &-0.5 &0.5 \nl
  $10^6$& 4&  25& 0.5&0.6 &0.58 &$4\times10^{-6}$  & -0.8&1.0 \nl
  $10^6$& 4&  50& 0.5&0.6 &0.29 &$1\times10^{-5}$  & -0.6&0.6 \nl
\enddata

 
\tablenotetext{a}{$t_{bind}$ is defined as the time at which 
$E_K+E_G+\delta E_B=0$}
\tablenotetext{b}{$t_{condense}$ is defined as the time at which 
$E_G/E_{\rm G, sheet}=1$ where $E_{\rm G, sheet}= -[(\nj^2 \pi^2/6) -1]$.}
\tablenotetext{c}{Evaluated at the crossing time 
$t_{cr}\equiv L/(2 E_{K,\ init})^{1/2}$}
\tablenotetext{d}{Volume-weighted average of $\log(\rho/\rb)$, evaluated at the
time when $E_K=E_{K,\, init}/2$ (see text).}
\tablenotetext{e}{Mass-weighted average of $\log(\rho/\rb)$, evaluated at the
time when $E_K=E_{K,\, init}/2$ (see text).}
\tablenotetext{\dagger}{Extrapolated} 
\tablenotetext{*}{Self-gravity has become unresolved at the grid scale 
before this measurement is made} 
\end{deluxetable}

\clearpage

\begin{figure}
\figurenum{1}
\plotone{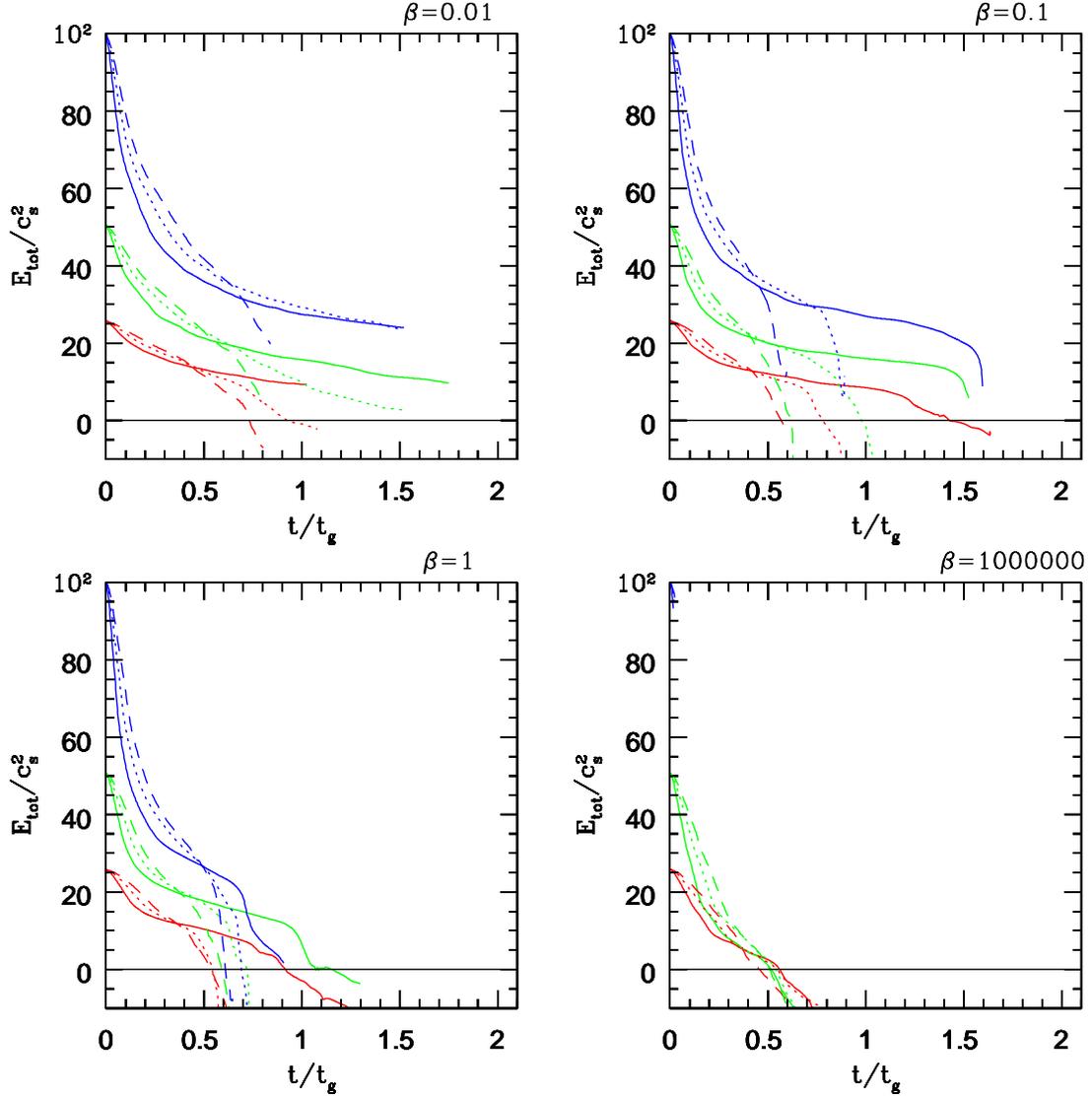}
\caption{Evolution of the total specific 
energy $E_{tot}/\cs^2\equiv (E_K + \delta E_B + E_G)/\cs^2$ for all models.  
Solid, dotted, and dashed curves have $n_J=2,3,$ and 4, respectively.}
\end{figure}

\clearpage

\begin{figure}
\figurenum{2}
\plotone{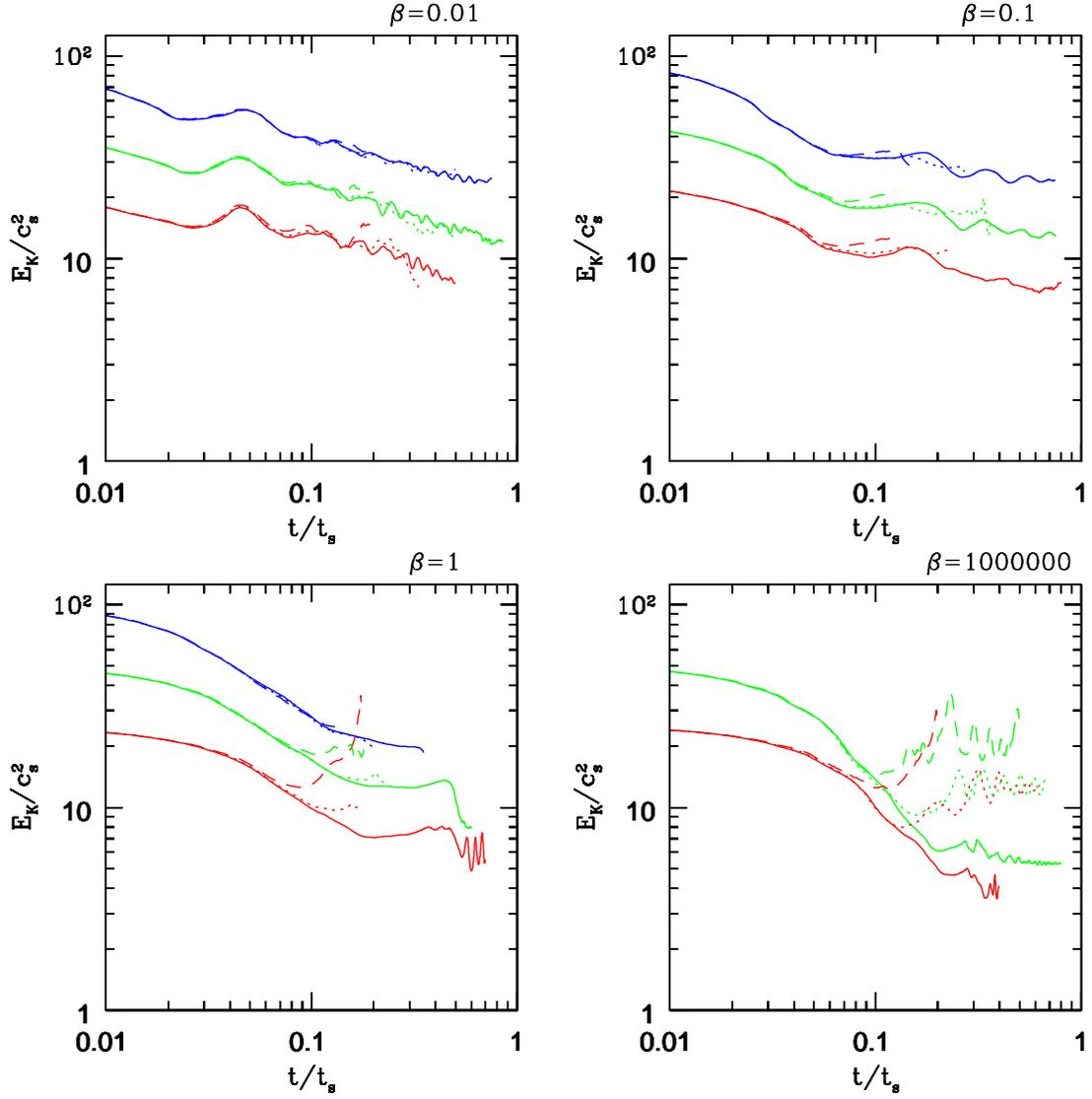}
\caption{Evolution of the specific kinetic
energy for all models.  Solid, dotted, and dashed curves have 
$n_J=2,3,$ and 4, respectively. }
\end{figure}

\clearpage
\begin{figure}
\figurenum{3}
\plotone{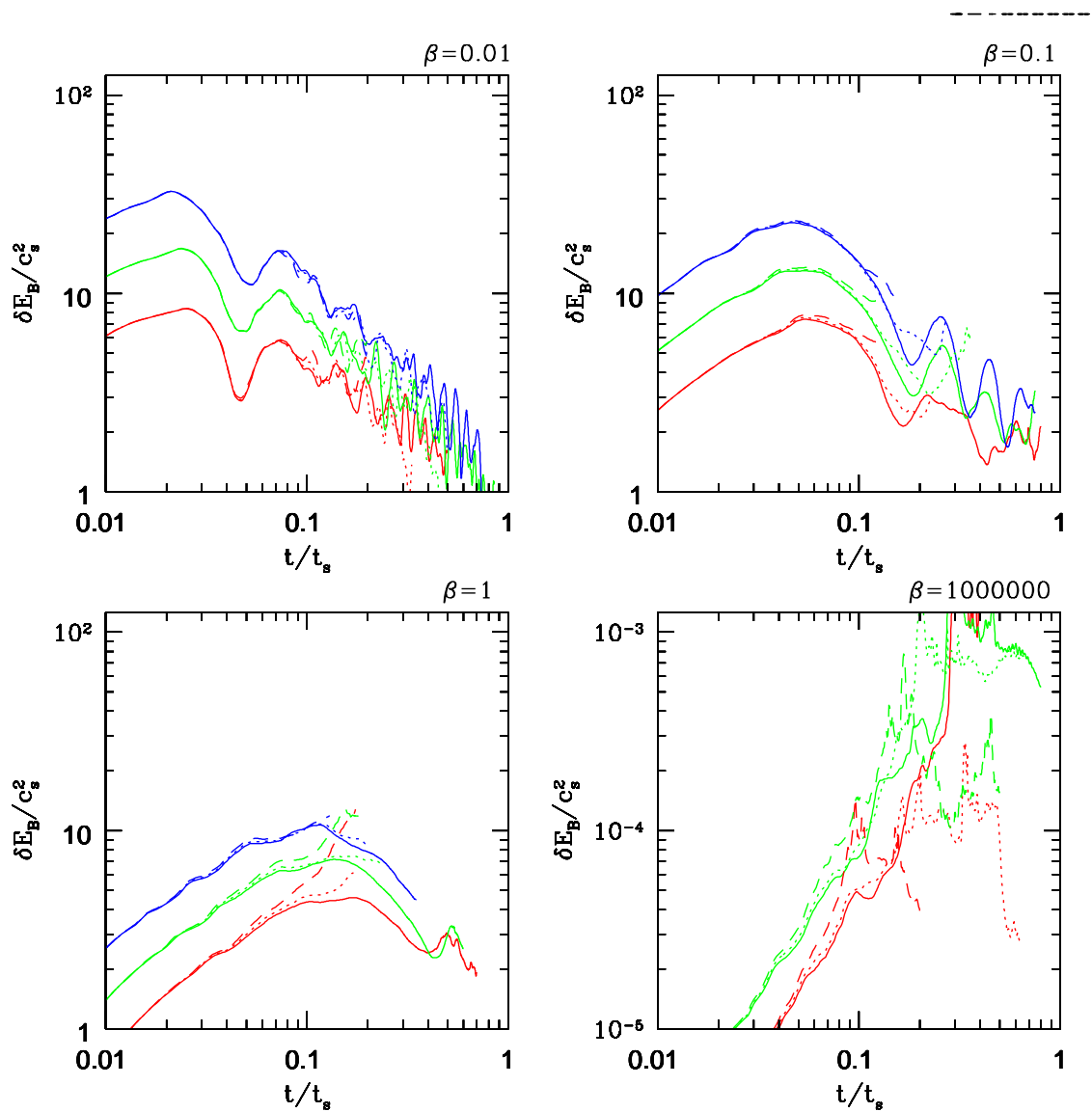}
\caption{Evolution of the perturbed magnetic energy (per unit
mass) for all models.  Solid, dotted, and dashed curves have 
$n_J=2,3,$ and 4, respectively.}
\end{figure}

\begin{figure}
\figurenum{4}
\plotone{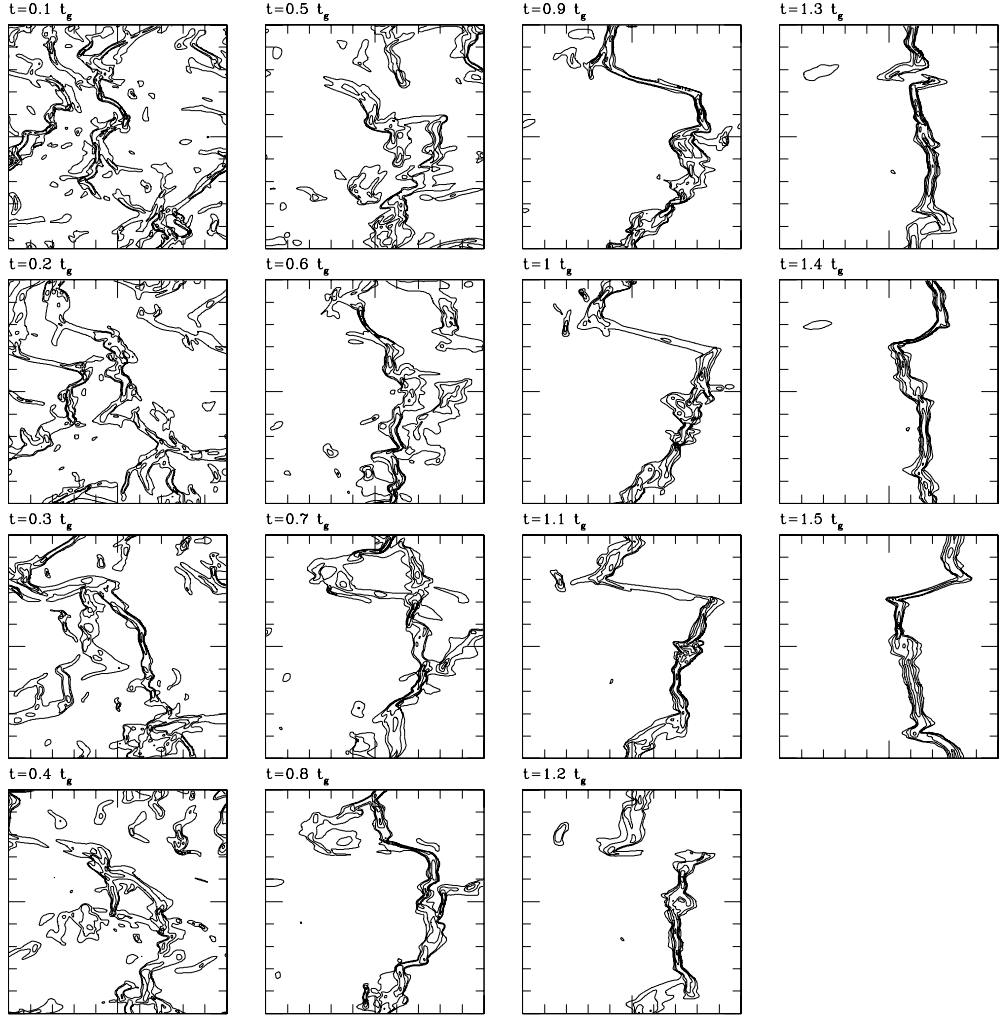}
\caption{Density snapshots of the $\beta=0.01$, $\nj=3$, $\ekinit=50$ model
cloud at temporal intervals of $0.1t_g$.  Density contours are at equal 
intervals in $\log(\rho/\rb)=0,0.5,1.,...$.  
}
\end{figure}

\begin{figure}
\figurenum{5}
\plotone{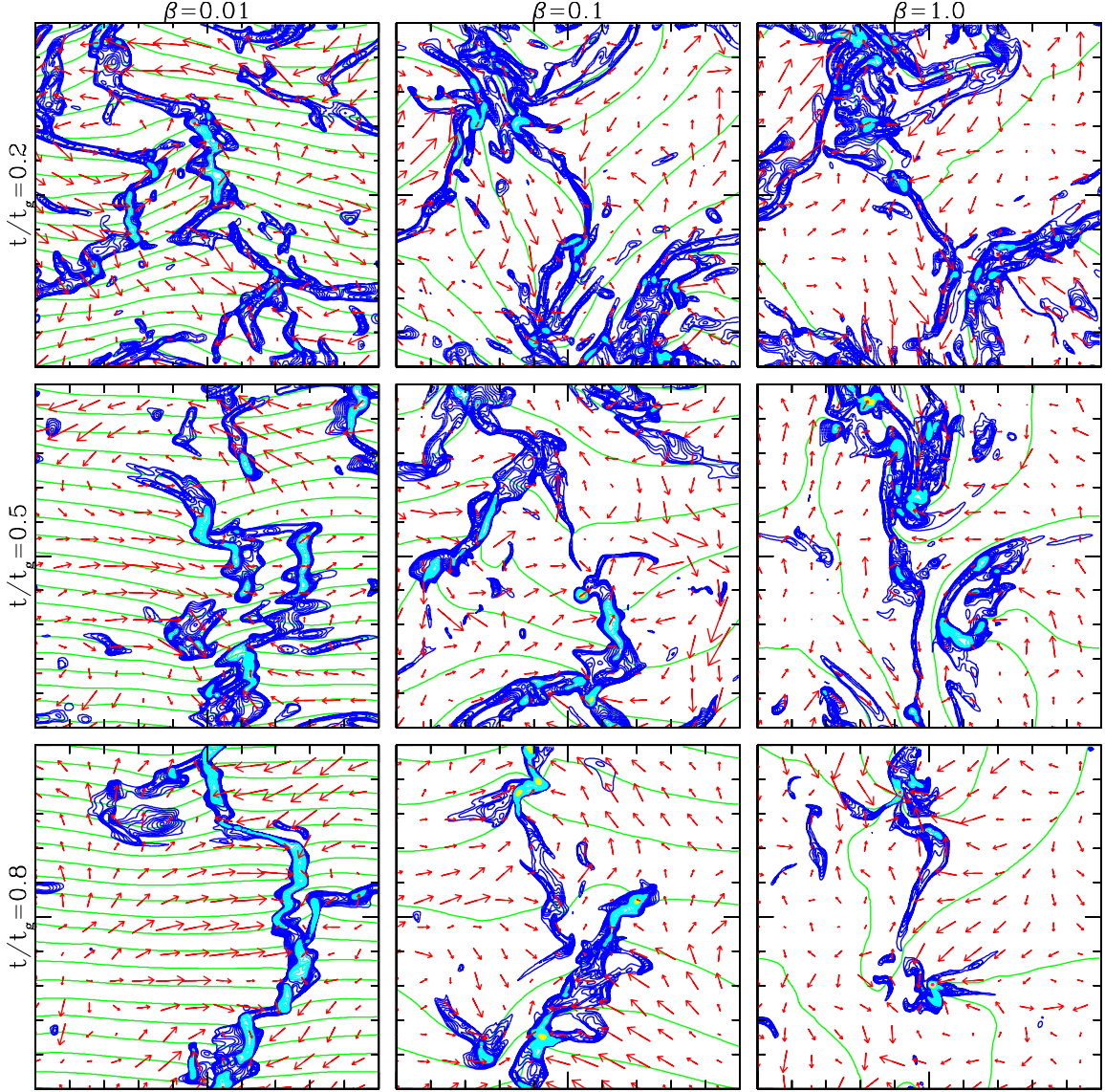}
\caption{Comparative evolution of cloud models with different magnetization.
Left, center, and right columns have $\beta=0.01$, 0.1, and 1, respectively.
$\ekinit=50$ and $\nj=3$ for all three runs.  Snapshots at times 
$t/t_g=0.2, 0.5, {\rm and} 0.8$ show density contours 
($\log(\rho/\rb)=$ 0-0.9:dark blue, 1-1.9:light blue, 2-2.9:yellow, 
3-5:magenta), velocity vectors (red), and magnetic field lines (green).
}
\end{figure}

\begin{figure}
\figurenum{6}
\plotone{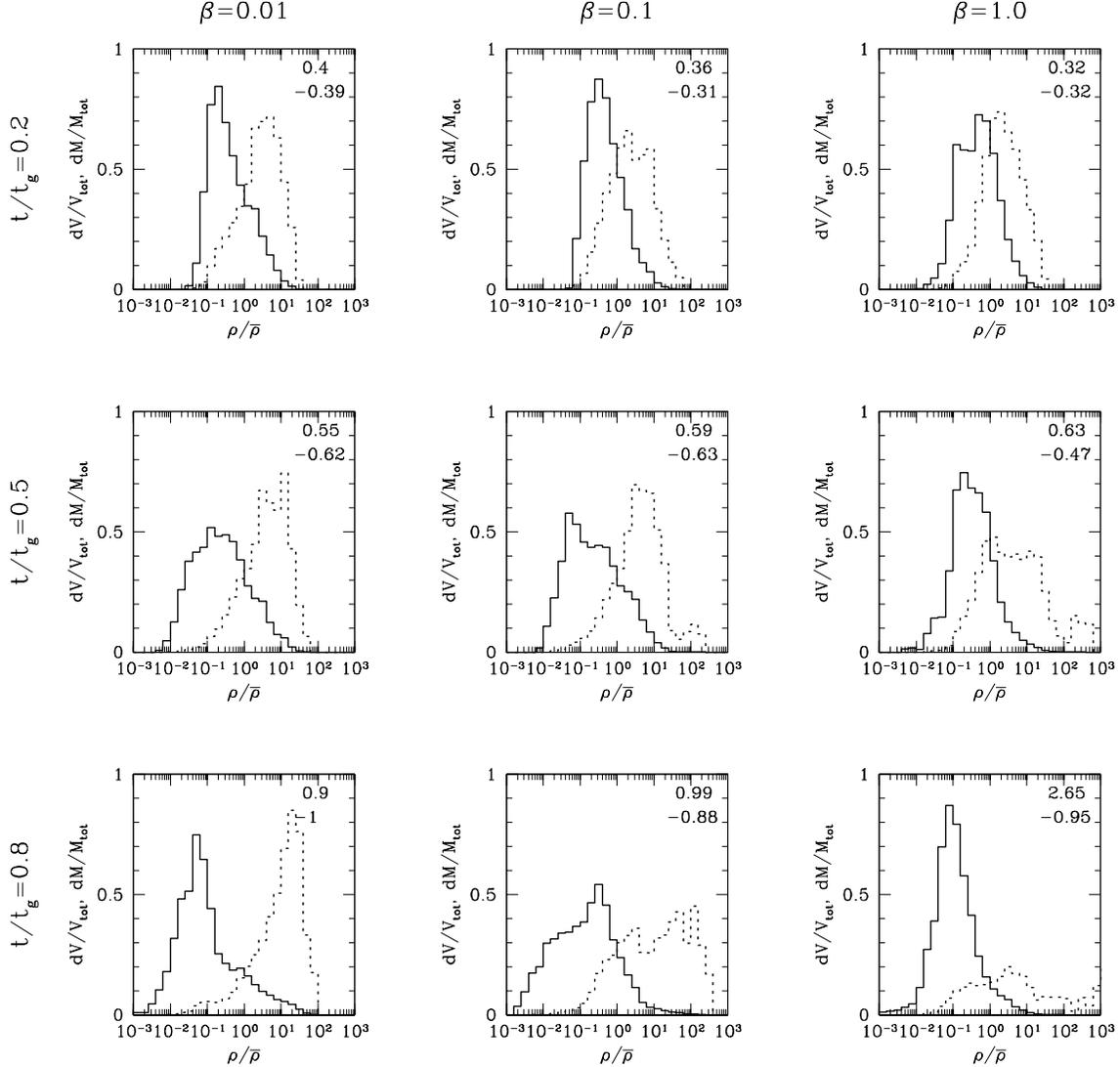}
\caption{Comparative evolution of density distributions in 
cloud models with different magnetization.  Same parameters and intervals as
in Fig. 5.  The solid histograms show the distribution of volume $dV/V_{tot}$,
and the dashed histograms show the distribution of mass $dM/M_{tot}$, as
a function of density, for each snapshot.  The numbers in the upper-right of
each panel indicate the values of the means 
$\langle \log(\rho/\rb)\rangle_M$ (top) and 
$\langle \log(\rho/\rb)\rangle_V$ (bottom) for each distribution.
}
\end{figure}

\begin{figure}
\figurenum{7}
\plotone{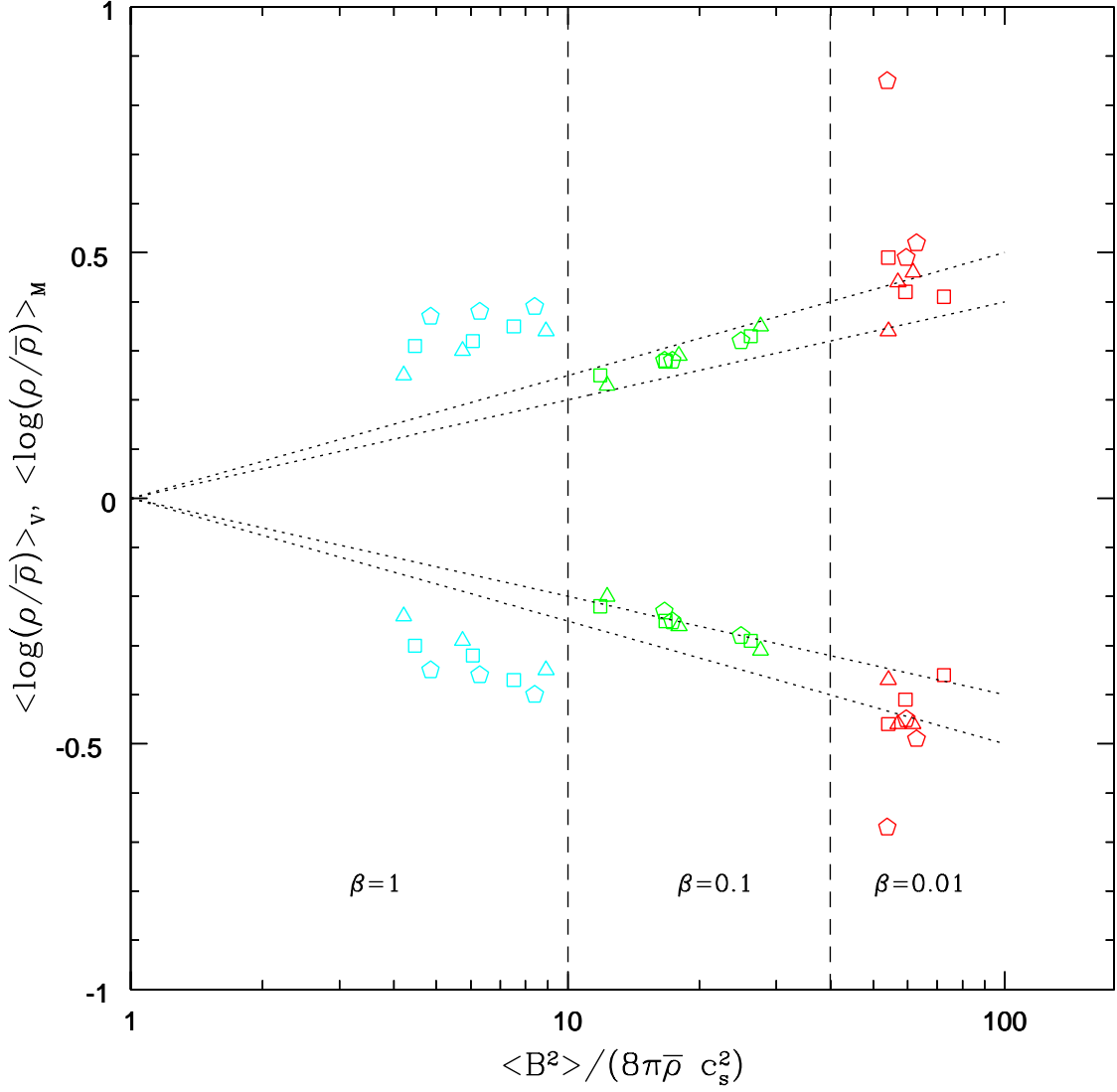}
\caption{Mass-averaged (upper points) and volume-averaged (lower points)
density contrast in model clouds when $E_K=0.5 \ekinit$, as a function of
total magnetic energy. Vertical dashed lines separate results from models with 
$\beta=1, 0.1, 0.01$.  Point type denotes Jeans number $\nj$: triangles, 
squares, pentagons for $\nj=2,3,4$, respectively.  The dotted lines indicate
logarithmic slopes of $\pm 0.2$ and $\pm 0.25$.
}

\end{figure}

\end{document}